\documentclass{iaus}
\usepackage{graphicx}

\newcommand{\degree}{\hbox{$^\circ$$$}}

\title[Low-latitude stellar substructure] 
{Mapping low-latitude stellar substructure with SEGUE photometry}

\author[J. T. A. de Jong et al.]   
{Jelte T. A. de Jong$^1$, Brian Yanny$^2$, Hans-Walter Rix$^1$, Eric
  F. Bell$^1$ and Andrew E. Dolphin$^3$}

\affiliation{$^1$Max-Planck-Institut f\"ur Astronomie, K\"onigstuhl
  17, 69117 Heidelberg, Germany \\ email: {\tt dejong@mpia.de} \\[\affilskip]
$^2$Fermi National Accelerator Laboratory, P.O. Box 500, Batavia, IL
  60510, United States \\[\affilskip]
$^3$Raytheon Corporation, 870 Winter Street, Waltham, MA 02451, United States
}

\pubyear{2008}
\volume{254}  
\pagerange{}
\jname{The Galaxy disk in cosmological context}
\editors{J. Andersen, J. Bland-Hawthorn \& B. Nordstr\"om, eds.}
\begin{document}

\maketitle

\begin{abstract}
Encircling the Milky Way at low latitudes, the Low Latitude Stream is
a large stellar structure, the origin of which is as yet unknown.
As part of the SEGUE survey, several photometric scans have been
obtained that cross the Galactic plane, spread over a longitude range
of 50\degree~ to 203\degree. These data allow a systematic study of
the structure of the Galaxy at low latitudes, where the Low Latitude
Stream resides. We apply colour-magnitude diagram fitting techniques
to map the stellar (sub)structure in these regions, enabling the
detection of overdensities with respect to smooth models. These
detections can be used to distinguish between different models of the
Low Latitude Stream, and help to shed light on the nature of the system.

\end{abstract}

\firstsection 
\section{Introduction}

During the past decade, the structure of the Milky Way (MW) has been
mapped in unprecedented detail, owing to deep, wide-field photometric
surveys, such as the Sloan Digital Sky Survey \cite[(SDSS, York et
al. 2000)]{york00} and 2MASS. Apart from an improved understanding of
the overall shape and stellar populations of the MW, these data have
unveiled a plethora of substructures on all scales. Two examples of
large structures are the stellar overdensity towards Canis Major
\cite[(CMa, Martin et al. 2004)]{martin04}, and the Low Latitude
Stream (LLS), or Monoceros stream, a ring-like structure that seems to
encircle the MW at low latitudes \cite[(Newberg et al. 2002, Ibata et
  al. 2003)]{newberg02,ibata03}.

The origin of these structures is as yet unclear, but several
hypotheses have been put forward. CMa might be (the remnant of) an
accreted dwarf galaxy \cite[(e.g. Martin et al. 2004,
Mart\'inez-Delgado et al. 2005, Bellazzini et al. 2006, Butler et
al. 2007, de Jong et
al. 2007)]{martin04,martinez05,bellazzini06,butler07,dejong07} with
the LLS being tidal debris stripped from CMa. \cite[Martin et
  al. (2005)]{martin05} and \cite[Pe\~narrubia et
al. (2005)]{penarrubia05} have shown that dynamical models of such an
accretion can indeed reproduce both the CMa overdensity as well as the
LLS.  On the other hand, as these structures are located at very low
Galactic latitudes, it cannot be ruled out that they are intrinsic to
the disk itself \cite[(e.g. Momany et al. 2006)]{momany06}. Another explanation for the LLS is that its stars
originate in the disk, but have been pulled out of the plane by
interactions with satellites \cite[(e.g. Kazantzidis et al. 2007,
Younger et al. 2008)]{kazantzidis07,younger08}.

\begin{figure}[t]
\begin{center}
\includegraphics[width=12cm]{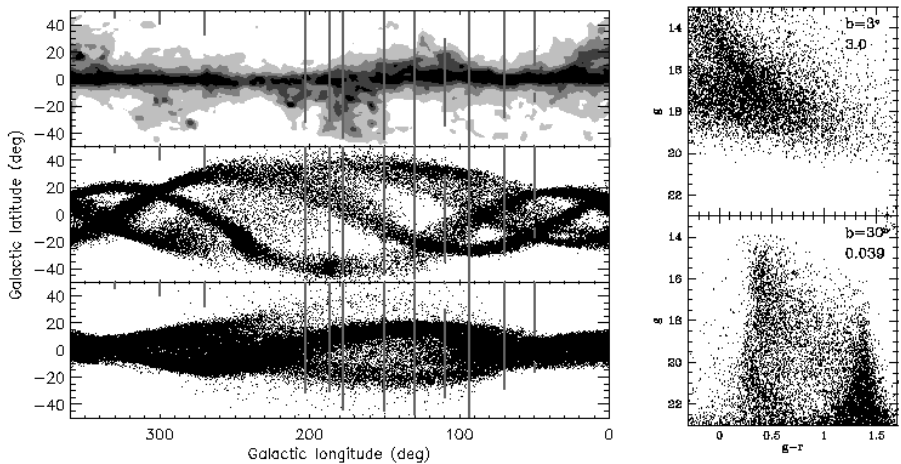}
\caption{
{\it Left:} Overview of imaging scans used for this analysis, indicated as
vertical lines, in Galactic coordinates. The top panel shows reddening according to the dust extinction maps from \cite[Schlegel et al. (1998)]{sfd}, with the
gray scale indicating regions with $E(B-V) >$0.1, 0.25, 0.5 and 1.0
mag. The middle and lower panels show the LLS models of \cite[Pe\~narrubia et
al. (2005)]{penarrubia05} and \cite[Martin et al. (2005)]{martin05},
respectively. {\it Right:} CMDs from the scan at $l$=94\degree. The
top CMD, at $b$=3\degree, is heavily extincted (E(B-V)=3 mag), while
the bottom CMD, at $b$=30\degree, is typical of the data used for our
CMD fitting analysis.}
\label{fig:dataoverview}
\end{center}
\end{figure}

Here we present preliminary results of the application of
colour-magnitude diagram (CMD) fitting techniques to photometry at low
Galactic latitudes taken as part of the SEGUE survey. These
techniques, developed in \cite[de Jong et al. (2008)]{sdssmatch},
allow us to map the 3-D distribution of stars. Fig.
\ref{fig:dataoverview} demonstrates that the resulting maps of
stellar (sub)structure along the SEGUE imaging scans can
help to distinguish between different models of the
LLS and provide crucial constraints for further modelling endeavours.

\section{Data and methods}

SEGUE (Sloan Extension for Galactic Understanding and Exploration), an
imaging and spectroscopic survey aimed at the study of the MW and its
stellar populations, is one of the constituent projects of the
extended SDSS survey (SDSS II). The photometric part of the SEGUE
survey consists of several 2.5\degree~ wide scans going through the
Galactic plane (see Fig. \ref{fig:dataoverview}), allowing a view of
the Galaxy at low latitudes. For the CMD-fitting analysis we restrict
ourselves to the two most sensitive bands, $g$ and $r$, which are
complete to $\sim$22nd magnitude. Although for the main survey the
photometric accuracy is at least 2\% down to these limits
\cite[(Ivezic et al. 2004)]{Iv04}, in crowded regions at low latitudes
the accuracy might be worse and the calibration in the data used here
is preliminary.  In Fig. \ref{fig:dataoverview} the coverage of the
scans between Galactic latitudes of +50\degree~ and -50\degree~ is
shown. Where possible the SEGUE scans were extended to high latitudes
by extracting 2.5\degree~ wide strips from SDSS data release 5
\cite[(Adelman-McCarthy et al. 2007)]{dr5}. We de-redden all data
using the dust maps from \cite[(Schlegel et al. 1998)]{sfd}, including
the correction suggested by \cite[Bonifacio et
al. (2000)]{bonifacio00}. Two CMDs from the scan at $l$=94\degree~ are
also shown in Fig. \ref{fig:dataoverview}. Very close to the plane,
the reddening is very high and de-reddening is unable to correct this
accurately. As poor de-reddening would limit the accuracy of our
results, we avoid regions with reddening higher than E(B-V)=0.2 mag.
The CMD in the bottom right of Fig. \ref{fig:dataoverview} is
representative of the the data used.

In CMD fitting, observed photometry is compared with models in order
to constrain the constituent stellar populations of stellar
systems. Traditionally, CMD fitting has been used mostly to determine
star formation histories and age-metallicity relations of isolated
objects, such as dwarf galaxies and globular clusters. We use the CMD
fitting software package MATCH \cite[(Dolphin 2001)]{match}, adapted
to solve for distance rather than age-metallicity evolution, to fit
the distance distribution of stars along the line-of-sight. A
demonstration and a detailed description of this technique and its
application to SDSS data can be found in \cite[de Jong et
  al. (2008)]{sdssmatch}. Here we shortly discuss the basic approach
used in the current analysis. 

Observed CMDs are fit with a set of model CMDs,
each of which corresponds to a certain age, metallicity and
distance. The models are created by populating
theoretical isochrones from \cite[Girardi et al. (2004)]{girardi04}
and convolving them with the SDSS photometric errors and completeness
(see de Jong et al. 2008). Using maximum-likelihood methods, MATCH
determines the best-fitting linear combination of model CMDs, thereby
providing the distribution of stars over age-metallicity-distance space.
We do this in two ways. First, fitting a narrow colour range of
0.3$<g$-$r<$0.8 and a single combination of age and metallicity
([Fe/H]=$-$0.8, 8 Gyr); this corresponds to measuring the density of
main-sequence stars as function of distance modulus. Second, we use a
wider colour range (0.1$<g$-$r<$0.8) that includes turn-off stars,
providing information on age and metallicity, and fit for three
different stellar populations: a `thick-disk' population with
[Fe/H]=$\sim-$0.8 and $t$=10 Gyr, a `halo' population with
[Fe/H]$\sim-$1.3 and $t$=13 Gyr, and a broad `general' population with
[Fe/H]=$\sim-$0.8 and 5$<t<$14 Gyr.

\section{Results}

\begin{figure}[t]
\begin{center}
\includegraphics[width=11.5cm]{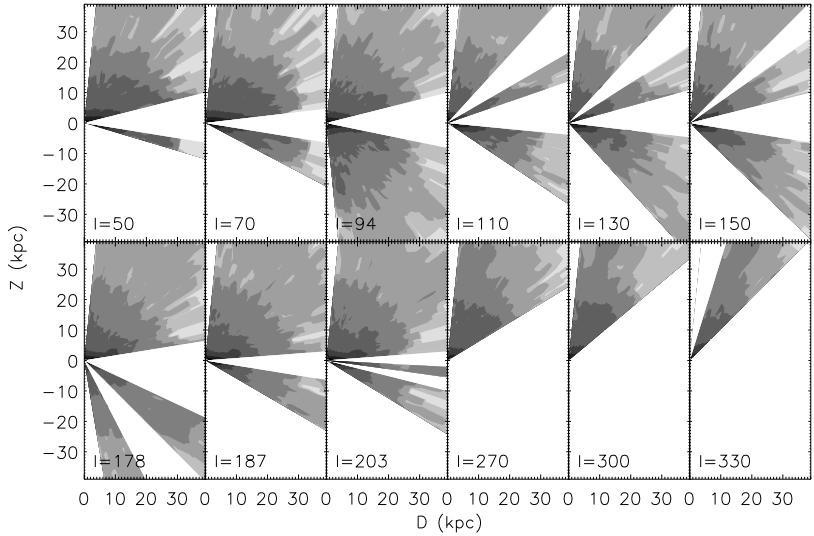}
\caption{
Stellar mass density resulting from the single population fits, as
function of distance from the sun along the Galactic plane, and height
above or below the plane, in kpc. Each panel shows a different imaging
scan, the Galactic longitude of which is listed at the bottom, and
darker gray levels correspond to higher densities.}
\label{fig:dens_1pop}
\end{center}
\end{figure}

\begin{figure}[t]
\begin{center}
\includegraphics[width=11.5cm]{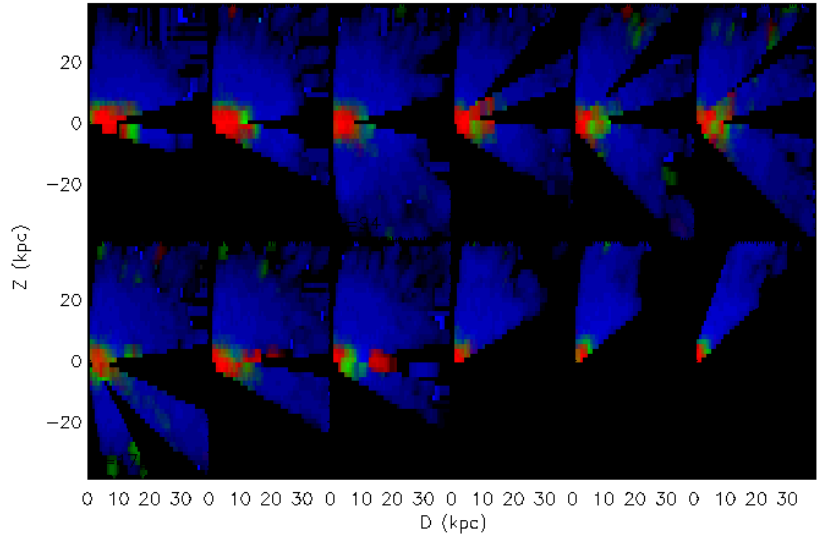}
\caption{ Distribution of stellar populations, following the fit
results for three different populations. Each panel is for a different
longitude, following the same lay-out as Fig. \ref{fig:dens_1pop}. The
thick-disk-like population with [Fe/H]$\simeq-$0.8 and $t$=10 Gyr is
colour-coded as red, the halo-like population with [Fe/H]$\simeq-$1.3 and
$t$=13 Gyr as blue, and the broad population with [Fe/H]$\simeq-$0.8
and 5$<t<$14 Gyr as green.  }
\label{fig:pops_realpop}
\end{center}
\end{figure}

From the fits described above we obtain the stellar mass in bins of
constant distance modulus. This can be converted to spatial stellar
mass density, giving contour maps such as the ones presented in
Fig. \ref{fig:dens_1pop} for the single population fits. As expected,
the density is clearly seen to decrease with increasing distance from
the plane and from the Galactic center. Fig. \ref{fig:pops_realpop} is
colour-coded with the type of population, based on the results from
the fits with the thick-disk-like (red), halo-like (blue) and very
broad (green) populations. It is clear that the thick disk and halo
are indeed fit with the appropriate populations, with the third, broad
population mostly visible in the regions where the halo and disk
populations have similar densities.

\begin{figure}
\begin{center}
\includegraphics[width=12cm]{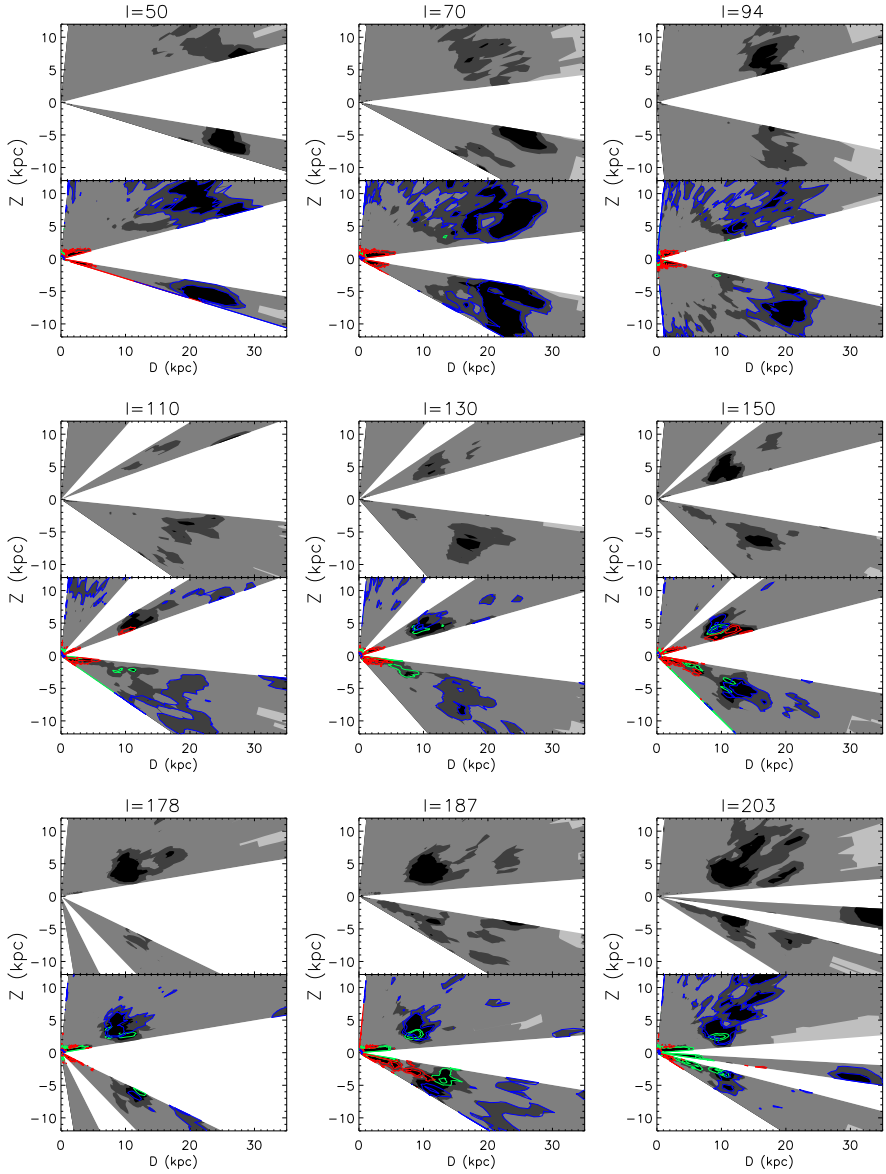}
\caption{
Overdensities at low Galactic latitudes. Residual maps for both single
population (upper subpanels) and triple population (lower subpanels)
fits for all stripes that cross the Galactic plane. Starting from
black, the gray scale levels correspond to areas with residual
densities $>$2, 2$>$1, 1$>-$1, and $<-$1 times the model
density. White areas contain no information due to an absence of data
or too high reddening. Coloured contours show overdensities of $>$2,
2$>$1 times the model density in the individual thick disk-like (red),
halo-like (blue) and broad (green) populations.}
\label{fig:subtracted}
\end{center}
\end{figure}

To increase the contrast of any substructures on top of the `smooth'
distribution of stars, a smooth model must be subtracted from the
stellar density maps. We assume a model with a double exponential thin
and thick disk and an axisymmetric power-law halo. Since the constraints
on the disk are limited due to a lack of data at the smallest
distances and the masking out of the lowest latitudes, we fix the thin
and thick disk scale lengths at 2.6 and 3.6 kpc,
respectively, and their scale heights at 0.3 and 1.0 kpc,
respectively. Using $R_\odot$=7.6 kpc (following \cite[Vall\'ee
(2008)]{vallee08}), our best-fit model gives a local thin disk density
of 0.08 $M_\odot$pc$^{-3}$, local thick disk and halo
normalisations of 0.058 and 0.0016, and an almost round ($q$=0.9) halo
with a power-law index of $-$3. These fit values agree well with
previous determinations \cite[(see e.g. Siegel et
  al. 2002)]{siegel02}, although our fit favours a rounder
halo than found in previous SDSS studies by \cite[
 Bell et al. (2008) and Juric et al. (2008)]{bell08,juric08}.
For the fits with three populations the model favours a lower thin
disk density and higher normalisations for the thick disk and
halo. This is due to the lack of a specific thin disk-like population
in the fits.
The residuals left after subtracting the best-fitting smooth models
from the stellar density maps are shown in Fig. \ref{fig:subtracted}.
In this figure we have zoomed in to regions close to the plane, where
the LLS is expected to be present.

\section{Discussion and Conclusions}

Colour-magnitude diagram fitting can successfully reproduce the distance
distribution of stars. This is corroborated by the fact that the disk
and halo model that best fits the density distributions in
Fig. \ref{fig:dens_1pop} gives densities and density ratios between
the different model components that are in good agreement with
previous determinations from star counts \cite[(e.g. Siegel et
al. 2002 and references therein)]{siegel02}.

After subtraction of a smooth Galaxy model, a wealth of substructure
becomes visible at low Galactic latitudes (Fig. \ref{fig:subtracted}).
When searching for pieces of the LLS, possible other overdensities
need to be taken into account. The scans at $l$=50\degree~ and
70\degree cross the Hercules-Aquila cloud \cite[(Belokurov et
  al. 2007)]{hercaqui}, a large overdensity extending both above and
below the plane to high latitudes at a distances between 10 and 20
kpc. Consequently, the extended overdensities seen in these scans in
Fig. \ref{fig:subtracted} are likely to be related to this structure
(it appears slightly more distant in our results because
Hercules-Aquila is more metal-poor than the populations used in our
fits). Furthermore, the Sagittarius stream lies partly along the scan
at $l$=203\degree, and is probably partly responsible for the
overdensities there. However, even with these restrictions, a large
number of overdensities is left, of which at least part corresponds to
the LLS. For example, the overdensities at D=10 kpc, Z=+5
kpc in the $l$=178\degree~ and $l$=187\degree~ scans correspond to the
original detection of the Monoceros overdensity by \cite[Newberg et
  al. (2002)]{newberg02}.

More accurate analysis of the distances and metallicities of the
detected overdensities is needed to piece together an overview of the
LLS and possible other substructures. One avenue of solving for the
degeneracies between distance, age, and metallicity is by using
colour-colour information to obtain metallicity estimates
\cite[(Ivezic et al. 2008)]{ivezic08}. With locations and
metallicities for the overdensities in place, a detailed picture of
the LLS will emerge that can help to shed light on the nature of this
structure.\\

Funding for the SDSS and SDSS-II has been provided by the Alfred
P. Sloan Foundation, the Participating Institutions, the National
Science Foundation, the U.S. Department of Energy, the National
Aeronautics and Space Administration, the Japanese Monbukagakusho, the
Max Planck Society, and the Higher Education Funding Council for
England. The SDSS Web Site is http://www.sdss.org/.

%


\begin{thebibliography}{}

\bibitem[Adelman-McCarthy et al.(2007)]{dr5}{Adelman-McCarthy et al.
  2007, \textit{ApJS}, 172, 634}
\bibitem[Bell et al.(2007)]{bell07}{Bell, E. F., et al.\ 2007, \textit{ApJ},
  680, 295}
\bibitem[Bellazzini et al.(2006)]{bellazzini06} Bellazzini, et al.\ 2006,
  \textit{MNRAS}, 366, 865
\bibitem[Belokurov et al.(2007)]{hercaqui}{ Belokurov, V., et al.\
  2007, \textit{ApJ}, 657, L89}
\bibitem[Bonifacio et al.(2000)]{bonifacio00} {Bonifacio, P., et al.\
  2000, \textit{AJ}, 120, 2065}
\bibitem[Butler et al.(2007)]{butler07}{Butler, D. J. et al.\ 2007, \textit{AJ}, 133, 2274}
\bibitem[de Jong et al.(2007)]{dejong07}{de Jong, J. T. A., et al.\ 2007,
  \textit{ApJ}, 662, 259} 
\bibitem[de Jong et al.(2008)]{sdssmatch}{de Jong, J. T. A., et al.\ 2008, \textit{AJ}, 135, 1361}
\bibitem[Dolphin(2001)]{match} Dolphin, A. E. 2001, \textit{MNRAS}, 332, 91
\bibitem[Girardi et al.(2004)]{girardi04} Girardi, L., et al.\ 2004, \textit{A\&A}, 422, 205
\bibitem[Ibata et al.(2003)]{ibata03}{Ibata, R. A., et al.\ 2003, \textit{MNRAS}, 340, L21}
\bibitem[Ivezi\'{c} et al.(2004)]{Iv04} Ivezi\'{c}, \v{Z}. et al.
2004, \textit{AN}, 325, 583
\bibitem[Ivezi\'{c} et al.(2008)]{ivezic08} {Ivezi\'{c}, \v{Z}. et al.
2008, \textit{ApJ}, subm., arXiv:0804.3850}
\bibitem[Juri\'{c} et al.(2008)]{juric08}{Juri\'{c}, M., et al.\ 2008,
  \textit{ApJ}, 673, 864}
\bibitem[Kazantzidis et al.(2007)]{kazantzidis07}{Kazantzidis, S., et al.\
  2007, \textit{ApJ}, subm. (arXiv:0708.1949)}
\bibitem[Martin et al.(2004)]{martin04}{Martin, N. F., et al.\ 2004,
  \textit{MNRAS}, 348, 12}
\bibitem[Martin et al.(2005)]{martin05} Martin, N. F., et al.\ 2005,
  \textit{MNRAS}, 362, 906
\bibitem[Martinez-Delgado et al.(2005)]{martinez05}
  Mart\'inez-Delgado, D., et al.\ 2005, \textit{ApJ}, 633, 205
\bibitem[Momany et al.(2006)]{momany06}{Momany, Y., et al.\ 2006,
  \textit{A\&A}, 451, 515}
\bibitem[Newberg et al.(2002)]{newberg02}{Newberg, H. J., et al.\ 2002, \textit{ApJ}, 569, 245}
\bibitem[Pe\~narrubia et al.(2005)]{penarrubia05} Pe\~narrubia, J.,
  et al.\ 2005, \textit{ApJ}, 626, 128
\bibitem[Schlegel et al.(1998)]{sfd} Schlegel, D.,
  Finkbeiner, D. \& Davis, M. 1998, \textit{ApJ}, 500, 525
\bibitem[Siegel et al.(2002)]{siegel02}{Siegel, M. H., et al.\ 2002,
  \textit{ApJ}, 578, 151}
\bibitem[York et al.(2000)]{york00} York, et al.\ 2000, \textit{AJ}, 120, 1579
\bibitem[Younger et al.(2008)]{younger08}{Younger, J. D., et al.\ 2008,
  \textit{ApJ}, 676, L21}
\end{thebibliography}
\end{document}